\documentstyle[12pt,amssymb]{article}
\begin{document}

\tolerance=5000

\def\pp{{\, \mid \hskip -1.5mm =}}
\def\cL{{\cal L}}
\def\be{\begin{equation}}
\def\ee{\end{equation}}
\def\bea{\begin{eqnarray}}
\def\eea{\end{eqnarray}}
\def\tr{{\rm tr}\, }
\def\nn{\nonumber \\}
\def\e{{\rm e}}

\begin{titlepage}

\begin{center}
\Large
{\bf Entropy and universality of Cardy-Verlinde formula in dark energy
universe}

\vfill

\normalsize

\large{ Iver Brevik$^\star$\footnote{Electronic
mail: iver.h.brevik@mtf.ntnu.no},
Shin'ichi Nojiri$^\spadesuit$\footnote{Electronic mail: nojiri@nda.ac.jp,
snojiri@yukawa.kyoto-u.ac.jp},
Sergei D. Odintsov$^{\heartsuit\clubsuit}$\footnote{Electronic mail:
odintsov@ieec.fcr.es Also at TSPU, Tomsk, Russia}},

and Luciano Vanzo$^\diamondsuit$\footnote{Electronic mail: vanzo@science.unitn.it}

\normalsize

\vfill

{\em $\star$ Department of Energy and Process Engineering, \\
Norwegian University of Science and Technology, \\
N-7491 Trondheim, NORWAY}

\

{\em $\spadesuit$ Department of Applied Physics,
National Defence Academy, \\
Hashirimizu Yokosuka 239-8686, JAPAN}

\

{\em $\heartsuit$ Institut d'Estudis Espacials de Catalunya (IEEC), \\
Edifici Nexus, Gran Capit\`a 2-4, 08034 Barcelona, SPAIN}

\

{\em $\clubsuit$ Instituci\`o Catalana de Recerca i Estudis
Avan\c{c}ats (ICREA), \\Barcelona, SPAIN}

\

{\em $\diamondsuit$ Dip.~di Fisica, Universit\`a di Trento 
and Ist.~Nazionale di Fisica Nucleare, \\
Gruppo Collegato di Trento, ITALY}

\end{center}

\vfill

\baselineskip=24pt
\begin{abstract}

We study the entropy of a FRW universe filled with dark energy
(cosmological constant, quintessence or phantom). For general or
time-dependent equation of state $p=w\rho$ the entropy is expressed in
terms of
energy, Casimir energy, and $w$. The
correspondent expression reminds one about 2d CFT entropy only
for conformal matter. At the same time, the cosmological Cardy-Verlinde
formula relating three typical FRW universe entropies remains to be
universal for any
type of matter. The same conclusions hold in modified gravity which
represents gravitational alternative for dark energy and which contains
terms growing at low curvature. It is interesting that BHs in modified
gravity are more entropic than in Einstein gravity. Finally, some
hydrodynamical examples
testing new shear viscosity bound, which is expected to be the consequence
of the holographic entropy bound, are presented for the early universe
in the plasma
era and for the Kasner metric.  It seems that the Kasner metric provides
a counterexample to the new shear viscosity bound.

\end{abstract}


\end{titlepage}

\section{Introduction}

There is growing evidence  from high redshift surveys of supernovae
and from WMAP data analysis that the current universe experiences a phase of
cosmic speed-up. The accepted explanation for this behavior is the
dominance of some dark energy contributing up to 70 percent of the critical
energy density. Nevertheless, it remains unclear what this dark energy is:
cosmological constant, quintessence, phantom, effective gravitational
contribution or something else. In the absence of a completely consistent
dark energy model a good strategy
would be to explore the general properties of FRW universe with dark
energy described as matter with a general (negative or
time-dependent) equation of state. Surprisingly, quite a lot of
information about the present and the future of such a universe may be obtained.

In particular, a number of issues related to entropy and energy of the
universe and their bounds may be understood. For instance, it seems clear
that the FRW equations are not so simple as they look as they may encode some
 quantum field theory structure via the holographic principle.
In a very interesting work \cite{EV}, a deep relation between the FRW
equations, conformal field theory entropy, and holography was established. First,
this work  proposed a holographical bound on the subextensive
entropy associated with Casimir energy. Second, it  showed
that the FRW universe entropy may be presented as a kind of Cardy entropy
in conformal field theory \cite{cardy}.
The corresponding expression is called the Cardy-Verlinde(CV)
formula. Moreover,
one more relation- the universal cosmological CV formula - may be
obtained
by rewriting the FRW equations in terms of three holographic entropies (or
energies). There is currently much activity in the study of various
aspects of the CV formula (see \cite{others} and references therein): its
holographic origin, the relation to the  brane-world
approach, and the description via AdS duals within the AdS/CFT set-up. It is also
remarkable that the CV formula should be generalized in the case of a general
(constant) equation of state \cite{Youm}, while the cosmological CV formula
remains  valid.

The purpose of the present work is to discuss the entropy,
Cardy-Verlinde-like
formulas, related consequences of holographic entropy bound for
(mainly) FRW
universe filled with dark energy where the effective equation of state is
negative or even time-dependent. In a similar fashion, these questions
are studied for modified gravity which represents a gravitational
alternative for dark energy. It is expected that a better understanding
of this topic may shed some light on questions about the origin
of holographic relations in the early universe as well as in the
current accelerating universe,
and on the origin of dark energy itself.

The paper is organized as follows. In the next section we discuss the
thermodynamic system which  corresponds to FRW universe
with a general equation of state which can be negative
(cosmological constant, phantoms or quintessence) or time-dependent.
The explicit expression for entropy of such FRW universe is found
and  is presented as a CV formula (in terms of energy and
Casimir energy). It is remarkable
that for a general equation of state, such a formula does not have simple
form
reminding about 2d CFT entropy. Another form of (cosmological)
CV formula (which is expected to have the holographic origin
and
which relates three different typical entropies of FRW universe) is
found to be  universal, like 2d CFT entropy. The entropy
bounds (including Bekenstein bound) for dark energy
universe and their dependence from critical radius are briefly mentioned.

Section three is devoted to the study of the same questions for modified
gravity which contains terms growing with the decrease of the curvature.
Such a theory describes the current accelerating universe and represents
the
gravitational alternative for dark energy. It is shown that the
cosmological
CV formula is universal, since it remains the same in both frames
(Jordan or
Einstein) used to describe such a gravity. In section four the black hole
thermodynamics for modified gravity is briefly discussed. It is shown
that for SAdS black holes the entropy is related to the area, with a
numerical coefficient different from the Einstein gravity case.
The relation of such an entropy to the CV formula is briefly mentioned.
Section five is related more to hydrodynamics and the early universe.
Namely, it was recently suggested some universal lower bound on the
relation between shear viscosity and entropy density. It is expected that
such bound directly follows from Bekenstein entropy bound. As shear
viscosity is typical for anisotropic universe we test the bound for
hydrodynamics or  Kasner universe. It seems that anisotropic
universe may give some counterexample for the bound.
Finally, summary and outlook are given in the last section.

\section{Thermodynamics of dark energy universe: energy and entropy}

Let us start from the simple thermodynamic system with the free energy
$F=F(V,T)$, where $V$ is  volume of the system and $T$
is  temperature. The pressure $p$, the energy density $\rho$, and the entropy
$S$ are given by
\be
\label{FI}
p=-{\partial F \over \partial V}\ ,\quad \rho ={1 \over V}\left( F - T {\partial F
\over \partial T}\right)\ ,\quad S=- {\partial F \over \partial T}\ .
\ee
The first law of the thermodynamics holds automatically:
\be
\label{FIb}
TdS=dE + pdV\ .
\ee
Here the total energy $E$ is given by $E=\rho V$.
The Boltzmann constant $k_B$ is chosen to be unity ($k_B=1$).
The free energy may be chosen in the following form:
\be
\label{FII}
F=-f_0 T^\alpha V^\beta\ ,
\ee
with some constants $f_0$, $\alpha$, and $\beta$. As a result
\be
\label{FIII}
p=\beta f_0 T^\alpha V^{\beta - 1}\ ,\quad
\rho= \left(\alpha - 1\right)f_0 T^\alpha V^{\beta - 1}\ ,\quad
S=\alpha f_0 T^{\alpha-1} V^\beta\ .
\ee
Defining a parameter $w$ by $p=w\rho$ (equation of state), we obtain
\be
\label{FIV}
w={\beta \over \alpha - 1}\ .
\ee
The case of interest is the negative equation of state, which is typical
for the current, dark energy,  universe.
The free energy can be rewritten as
\be
\label{FF1}
F=-f_0 T\left(T^{1 \over w} V\right)^\beta\ ,
\ee
which tells that the general free energy of the matter with $w$ has the following form
\be
\label{FF2}
F_w(T,V)=T \hat F\left(T^{1 \over w}V\right)\ .
\ee
Here $\hat F(x)$ is a function depending on the matter.

For $\alpha=4$ and $\beta=1$,  the classical radiation in
4-dimensional spacetime is restored:
\be
\label{FV}
p=f_0 T^4\ ,\quad \rho = 3 f_0 T^4\ .
\ee
In order to obtain ideal gas,  the free energy should look as
\be
\label{FVI}
F=f_0\left(T^\alpha V^\beta - T\right)\ .
\ee
It is interesting that the last term does not contribute to $p$ ( $\rho$)
but does contribute to
the entropy $S$. In the limit that $\alpha\to 1$ and $\beta\to 0$ with
finite $c_1=f_0\left(\alpha - 1\right)$ and $c_2=\beta f_0$, we obtain
\be
\label{FVII}
F=T\ln \left(T^{c_1} V^{c_2}\right)\ ,\quad p = c_2 TV\ ,\quad \rho=c_1 TV\ .
\ee
Then $c_2$ can be identified with the number $N$ of the molecules in the gas $c_2=N$
and $c_1={3 \over 2}N$ for the monoatomic molecule.
One can also obtain dust by choosing $\beta=0$:
\be
\label{FVIII}
p=0\ ,\quad \rho =\left(\alpha - 1\right)f_0 T^\alpha V^{-1}\ .
\ee

We may consider the case that the entropy is constant $S=S_0$
which is typical for adiabatically expanding universe where first law
of thermodynamics holds. From
(\ref{FIII}) it follows
\be
\label{FIX}
T=\left(\alpha f_0\right)^{-{1 \over \alpha -1}}S_0^{1 \over \alpha - 1} V^{-w}\ .
\ee
Here $w$ is given in (\ref{FIV}).

Let us apply the above considerations to the
 $(n+1)$-dimensional FRW metric of the form:
\be
\label{I}
ds^2=g_{\mu\nu}dx^\mu dx^\nu = -d\tau^2 + a^2(\tau) \gamma_{ij}dx^i dx^j\ ,
\ee
where the $n$-dimensional metric $\gamma_{ij}$ is parametrized by $k=-1,0,1$.
In the following,  mainly  the $k=1$ case is considered.
Since $V=a^n \int d^n x \sqrt\gamma $, the temperature of the universe is
\be
\label{FX}
T \propto a^{-nw}\ .
\ee
By combining (\ref{FIII}) and (\ref{FIX}), the total energy
$E=\rho V$ is given by
\be
\label{FXI}
E=\left(\alpha - 1\right) \alpha^{-{\alpha \over \alpha -1}} f_0^{-{\alpha \over \alpha - 1}}
S_0^{\alpha \over \alpha - 1} V^{-w} \propto a^{-nw}\ .
\ee
The  $a$-dependence in (\ref{FX}) and (\ref{FXI}) reproduces the
corresponding results in
\cite{Youm}.

Rescaling the entropy and the volume as $S_0\to \lambda S_0$ and $V\to
\lambda V$,
from the expression  (\ref{FXI}), we obtain
\be
\label{FXII}
E\to \lambda^{{1 \over \alpha - 1} + 1 - w}E\ .
\ee
If the energy is extensive,  $E\to \lambda E$. For the extensive part of
of the energy it follows
\be
\label{FXIII}
\alpha = 1 + {1 \over w}\ ,\quad \beta = 1\ .
\ee
In order to obtain the expression of $\beta$ in (\ref{FXIII}),
 Eq.(\ref{FIV}) should be used.

The following free energy for general equation of state may be considered:
\be
\label{FXIV}
F=-f_0 T^{1+{1 \over w}}V\left(1 + f_1 T^{-{2 \over nw}}V^{-{2 \over n}}\right)\ .
\ee
If there is no the second term, the first term gives the extensive energy.
Note that  $p=w\rho$ even if the second term is included.
As a result, the energy and entropy of the thermal universe follow
\bea
\label{FXV}
E&=&{f_0 \over w}T^{1+{1 \over w}}V\left(1 +
\left(1-{2 \over n}\right)f_1 T^{-{2 \over nw}}V^{-{2 \over n}}\right)\ ,\nn
S&=&f_0 T^{1 \over w}V\left(\left(1 + {1 \over w}\right)
+ \left(1 + {1 \over w} - {2 \over nw}\right)f_1 T^{-{2 \over nw}}V^{-{2 \over n}}\right)\ .
\eea
As clear from (\ref{FI}) and (\ref{FII}), the entropy becomes negative
(unphysical case) if  $f_0$ or
$\alpha$ are chosen to be negative. If the terms containing $f_1$ can be neglected,
as clear from the
Eqs. (\ref{FXV}),  the entropy $S$ becomes negative if
\begin{enumerate}
\item $f_0<0$ and $w<-1$ : in this case, the energy $E$ is positive.
\item $f_0>0$ and $0>w>-1$ : in this case, $E$ also becomes negative.
\end{enumerate}
We should also note the energy (if we neglect the terms containing $f_1$) is positive
(negative) if $f_0$ is positive (negative).
The case that the entropy is negative would be unphysical and should be excluded.
Then the case that $w<-1$ and the energy $E$ is positive,  and the case that $0>w>-1$
and the energy $E$ is negative,  should be excluded.

The sub-extensive part of the energy $E_C$, which is called the Casimir
energy, is given by
\be
\label{FXVI}
E_C = n\left(E+pV-TS\right)=-nV^2{\partial \over \partial V}\left({F \over V}\right)
= -2 f_0f_1 T^{1+{1 \over w} - {2 \over nw}}V^{1 - {2 \over n}}\ .
\ee
The extensive part of the energy $E_E$ has the following form:
\be
\label{FXVII}
E_E=E-{1 \over 2}E_C={f_0 \over w} T^{1+{1 \over w}}V\left(1 +
\left(1-{2 \over n} + w\right)f_1 T^{-{2 \over nw}}V^{-{2 \over n}}\right)\ .
\ee
 From the last expression in (\ref{FXV}), we obtain
\be
\label{FXVIII}
T \sim S^w \left(1 + \left(1 + {1 \over w}\right)^{-1}\left(1 -{2 \over nw} + {1 \over w}
\right)f_1 T^{-{2 \over nw}}V^{-{2 \over n}}\right)\ ,
\ee
and
\be
\label{FXIX}
E_E \sim S^{w+1} V^{-w} + {\cal O}\left(f_1^2\right)\ ,\quad
E_C \sim S^{w+1 - {2 \over n}} V^{-w} + {\cal O}\left(f_1^2\right)\ ,
\ee
which reproduce the behaviors in \cite{Youm}.
When the size of the universe is large, the second terms in $S$  (\ref{FXV}) and in $E_E$
 (\ref{FXVII}) are sub-dominant and we obtain
\be
\label{FXIXb}
S\sim f_0 T^{1 \over w}V\left(1 + {1 \over w}\right) \ ,\quad
E_E \sim {f_0 \over w} T^{1+{1 \over w}}V\ .
\ee
Then  combining (\ref{FXVI}) and (\ref{FXIXb}), for the FRW metric (\ref{I}) with $k=1$,
one gets
\bea
\label{FXIXc}
S&\sim& f_0\left(1+{1 \over w}\right)\left(-{2f_0^2 f_1 \over n}
\right)^{-{n \over 2\left(\left(w+1\right)n - 1\right)}}
V_0^{wn \over (w+1)n -1 }
\left[a^{nw} \sqrt{E_E E_C}\right]^{n \over (w+1)n-1} \nn
&=& A\left[a^{nw} \sqrt{\left(2E-E_C\right)E_C}\right]^{n \over (w+1)n-1} \nn
A&\equiv& f_0\left(1+{1 \over w}\right)\left(-{4f_0^2 f_1 \over n}
\right)^{-{n \over 2\left(\left(w+1\right)n - 1\right)}}
V_0^{wn \over (w+1)n -1 }\ .
\eea
Here $V_0=\int d^n x \sqrt{\gamma}$. Eq.(\ref{FXIXc}) reproduces Eq.(20) in \cite{Youm}
if we identify
\be
\label{FXIXd}
A=\left({2\pi \over \sqrt{\alpha\beta}}\right)^{n \over (w+1)n - 1}\ .
\ee
This expression represents one of the forms of Cardy-Verlinde formula
\cite{EV} for general equation of state.

As there are astrophysical indications that dark energy currently
dominates at the thermal universe our main interest is related with the
case  where $w$ can be negative.
One usually denotes the matter as quintessence if $-{1 \over 3}>w>-1$ and
as phantom \cite{phantom} if $w<-1$.
When $w=-1$, the situation corresponds  to the cosmological constant.
First we should note that entropy $S$ (\ref{FXIXc}) becomes singular at
$w=-1 + {1 \over n}$, which occurs since the product $E_C E_E$ becomes independent of the
temperature. If the entropy $S$ is conserved, Eq.(\ref{FXIXc}) indicates
that
the product $E_C E_E$ increases if the size of the universe $a$ increases when $w$
is negative. The entropy may be conserved but we may consider the variation of
the entropy as a change of the initial condition.

When $0>w>-1 + {1 \over n}$, if we keep $E_C E_E$ to be constant,
 Eq.(\ref{FXIXc}) shows
that $S$ decreases if $a$ increases. When $w<-1 + {1 \over n}$,
 $S$ increases if $a$ increases
but $S$ decreases if $E_C E_E$ increases. As is seen from (\ref{FXV}), the
specific heat
${dE \over dT}$ with fixed volume ($V$ is a constant) becomes negative, when $0>w>-1$.
For the phantom ($w<-1$), the specific heat is positive and for the cosmological constant,
the specific heat vanishes.

For the current realistic universe the case that there are many kinds of
matter (with dark energy dominance ) is typical. In such a case the free
energy may be
written as sum over various contributions
\be
\label{FXX}
F=-\sum_i f_{i0} T^{1+{1 \over w_i}}V\left(1
+ f_{i1} T^{-{2 \over nw_i}}V^{-{2 \over n}}\right)\ .
\ee
Then one gets
\bea
\label{FXXI}
E&=&\sum_i{f_{i0} \over w_i}T^{1+{1 \over w_i}}V\left(1 +
\left(1-{2 \over n}\right)f_{i1} T^{-{2 \over nw_i}}V^{-{2 \over n}}\right)\ ,\nn
S&=&\sum_i f_{i0} T^{1 \over w_i}V\left(\left(1 + {1 \over w_i}\right)
+ \left(1 + {1 \over w_i} - {2 \over nw_i}\right)f_{i1} T^{-{2 \over nw_i}}
V^{-{2 \over n}}\right)\ ,\nn
E_C &=& -2 \sum_i f_{i0} f_{i1} T^{1+{1 \over w_i} - {2 \over nw_i}}V^{1 - {2 \over n}}\ ,\nn
E_E &=& \sum_i{f_{i0} \over w_i} T^{1+{1 \over w_i}}V\left(1 +
\left(1-{2 \over n} + w_i\right)f_{i1} T^{-{2 \over nw_i}}V^{-{2 \over n}}\right)\ .
\eea
Thus, in case that there are several types of matter, we cannot obtain a
simple relation
(\ref{FXIXc}). Nevertheless, an inequality follows
\be
\label{FXXII}
S \geq S_i \sim
A_i \left[a^{nw_i} \sqrt{\left(2E_i-E_{Ci}\right)E_{Ci}}\right]^{n \over (w_i+1)n-1}\ .
\ee
Here
\be
\label{FXXIII}
A_i\equiv f_{0i}\left(1+{1 \over w_i}\right)\left(-{4f_{0i}^2 f_{1i} \over n}
\right)^{-{n \over 2\left(\left(w_i+1\right)n - 1\right)}}
V_0^{w_in \over \left(w_i+1\right)n -1 }\ .
\ee
As $S=\sum_i S_i$ and $S_i\geq 0$, the inequality (\ref{FXXII}) holds for arbitrary $i$.
With the entropy $S$ (\ref{FXXI}), at  high temperature the
matter with small
and positive $w_i$ dominates. We now denote the quantities related with
the matter with smallest
but positive $w_i$ by the index ``min''. At high temperature, instead of
(\ref{FXIXc}), one gets
\be
\label{FXXIV}
S \sim A_{\rm min}\left[a^{nw_{\rm min}} \sqrt{\left(2E_{\rm min}-E_{C\,{\rm min}}
\right)E_{C\,{\rm min}}}\right]^{n \over (w_{\rm min}+1)n-1} \ .
\ee
On the other hand, at low temperature as in current universe, if
all the $w_i$'s are positive, the matter with large $w_i$ dominates. We now denote
the quantities related with the matter for largest $w_i$ by the index
``max''.
Then at low temperature
\be
\label{FXXV}
S \sim A_{\rm max}\left[a^{nw_{\rm max}} \sqrt{\left(2E_{\rm max}-E_{C\,{\rm max}}
\right)E_{C\,{\rm max}}}\right]^{n \over (w_{\rm max}+1)n-1} \ .
\ee
If there is a dark energy (say, phantom) with negative $w$, such a matter
dominates at low
temperature
\be
\label{FXXVb}
S \sim A_p\left[a^{nw_p} \sqrt{\left(2E_p-E_{pC}
\right)E_{pC}}\right]^{n \over (w_p+1)n-1} \ .
\ee
Here we have denoted the quantities related with the phantom
matter by
the index ``$p$''.
Note that for negative equation of state the above universe entropy
formula does not remind one about the well-known Cardy formula in CFT.
Since the entropy is given by
\be
\label{P1}
S\sim f_{p0} T^{1 \over w_p}V\left(\left(1 + {1 \over w_p}\right)
+ \left(1 + {1 \over w_p} - {2 \over nw_p}\right)f_{p1} \left(T^{1 \over w_p}
V\right)^{-{2 \over n}}\right)\
\ee
for conserved entropy, $T^{1 \over w_p}V$ is a constant:
\be
\label{PFXXIX}
T^{1 \over w_p}V = C\ .
\ee
Then  the energy $E$ can be rewritten as
\bea
\label{PFXXX}
E &\sim& {f_{p0} \over w_p}C T \left(1 +
\left(1-{2 \over n}\right)f_{p1} C^{-{2 \over n}}\right)\ ,\nn
&=&  {f_{p0} \over w_p}C^{w_p + 1}V^{-w_p} \left(1 +
\left(1-{2 \over n}\right)f_{p1} C^{-{2 \over n}}\right)\ ,\nn
&=& {f_{p0} \over w_p}C^{w_p + 1}V_0^{-w_p}a^{-nw_p} \left(1 +
\left(1-{2 \over n}\right)f_{p1} C^{-{2 \over n}}\right)\ .
\eea
Thus, the energy is linear with the temperature. In the last line, we have
considered
the FRW metric (\ref{I}).
Generally in the FRW metric, if we have the relation $p=w\rho$, we find
$\rho \propto a^{-n\left(1+w\right)}$ (energy conservation) and $E=\rho V
\propto a^{-nw}$, which
is consistent with (\ref{PFXXX})\ .
If there is only dark matter with $w<0$ in the universe,
the relation (\ref{PFXXIX}) is valid even at high temperature.
When the universe expands and the radius grows, the temperature
grows too
 and also the energy $E$ and the energy density $\rho$ behave as
$E\sim a^{-nw_p}$ and $\rho\sim a^{-n\left(w_p + 1\right)}$.
As a result the density becomes
large and might generate some future singularities (like Big Rip).

As an example the system with dust and quintessence or phantom, where $w$
is negative, may be considered.
If we assume that there is no internal structure in the dust, the energy of the
dust does not depend on the temperature and the free energy, corresponding to
(\ref{FVIII}), becomes a constant:
$F=E_{D0}$. Then the total free energy can be assumed to be given by
\be
\label{FXXVI}
F=E_{D0} - f_{p0} T^{1+{1 \over w_p}}V\left(1 + f_{p1}
T^{-{2 \over nw_p}}V^{-{2 \over n}}\right)\ .
\ee
Thus, one obtains
\bea
\label{FXXVII}
E&=& E_{D0} + {f_{p0} \over w_p}T^{1+{1 \over w_p}}V\left(1 +
\left(1-{2 \over n}\right)f_{p1} \left(T^{-{1 \over w_p}}V\right)^{-{2 \over n}}\right)\ ,\nn
S&=&f_{p0} T^{1 \over w_p}V\left(\left(1 + {1 \over w_p}\right)
+ \left(1 + {1 \over w_p} - {2 \over nw_p}\right)f_{p1} \left(T^{1 \over w_p}
V\right)^{-{2 \over n}}\right)\ .
\eea
Note that  dust does not contribute to the entropy.
The energy of the dust is not extensive nor sub-extensive.
The extensive and sub-extensive (Casimir) parts of the energy of the phantom or
quintessence are given by
\bea
\label{FXXVIII}
E_{pC} &=& -2 f_{p0} f_{p1} T^{1+{1 \over w_p} - {2 \over nw_p}}V^{1 - {2 \over n}}\ ,\nn
E_{pE} &=& {f_{p0} \over w_p} T^{1+{1 \over w_p}}V\left(1 +
\left(1-{2 \over n} + w_p\right)f_{p1} T^{-{2 \over nw_p}}V^{-{2 \over n}}\right)\ .
\eea
If we assume the entropy $S$ is conserved, from the expression of $S$ (\ref{FXXVII}),
we find $T^{1 \over w_p}V$ is a constant:
\be
\label{FXXIX}
T^{1 \over w_p}V = C\ .
\ee
Then the energy $E$  (\ref{FXXVII}) can be rewritten as
\bea
\label{FXXX}
E &=& E_{D0} + {f_{p0} \over w_p}C T \left(1 +
\left(1-{2 \over n}\right)f_{p1} C^{-{2 \over n}}\right)\ ,\nn
&=& E_{D0} + {f_{p0} \over w_p}C^{w_p + 1}V^{-w_p} \left(1 +
\left(1-{2 \over n}\right)f_{p1} C^{-{2 \over n}}\right)\ ,\nn
&=& E_{D0} + {f_{p0} \over w_p}C^{w_p + 1}V_0^{-w_p}a^{-nw_p} \left(1 +
\left(1-{2 \over n}\right)f_{p1} C^{-{2 \over n}}\right)\ .
\eea
Then energy is again linear in the temperature. In the last line, we have
considered
the FRW metric (\ref{I}).
Generally in the FRW metric, if we have the relation $p=w\rho$, we find
$\rho \propto a^{-n\left(1+w\right)}$ and $E=\rho V \propto a^{-nw}$, which
is consistent with the last expression for the phantom or quintessence in (\ref{FXXX})\ .

Taking into account the recent cosmological considerations
of variations of fundamental constants, one may start from
 the case that $w_p$ depends on the time $t$. Of course,
this may be negative (or sign-changing) function.
The energy conservation condition looks like
\be
\label{WT1}
0=\dot \rho_p + n{\dot a \over a}\left(\rho_p + p_p\right)\ ,
\ee
by assuming $\rho_p=w_p(t)p_p$. The following expression may be found
\be
\label{WT2}
\rho_p =a^{-n\left(1+w_p(t)\right)} \e^{n\int^t \dot w_p(t')\ln a(t')dt'}\ .
\ee
The energy in such a universe is
\be
\label{WT3}
E_p =\rho_p V = a^{-n w_p(t)} \e^{n\int^t \dot w_p(t')\ln a(t')dt'}V_0\ .
\ee
If the spacetime expansion is adiabatic and thermodynamical quantities can
be defined,
Eqs. (\ref{FI}) are valid. Thus, if we define a free energy as in the
phantom part of Eq.(\ref{FXXVI}), we can obtain  the entropy and energy
as in (\ref{FXXVII}) and the extensive and sub-extensive parts of the energy as in
(\ref{FXXVIII}).
Then if we define a variable $\xi$ by
\be
\label{WT4}
T=V^{-w_p(t)}\xi\ ,
\ee
extracting the phantom part $E_p$ from the expression of $E$ in (\ref{FXXVII}),
we obtain
\be
\label{WT5}
E_p={f_{p0} \over w_p(t)}a^{-nw_p(t)} V_0^{-w_p(t)} \xi^{1 + {1 \over w_p(t)}}
\left(1 + \left(1 - {2 \over n}\right)f_1 \xi^{-{2 \over nw_p(t)}}\right)\ .
\ee
By comparing (\ref{WT3}) with (\ref{WT5}), one finds
\bea
\label{WT6}
\xi &=&\left({w_p(t) \over f_{p0}}\right)^{w_p(t) \over w_p(t) + 1}V_0^{w_p(t)}
\e^{{nw_p(t) \over w_p(t) + 1}\int^t \dot w_p(t')\ln a(t') dt'} \nn
&& \times\left\{
1 - f_1{1 - {2 \over n} \over 1+{1 \over w_p(t)}}
\left({w_p(t) \over f_{p0}}\right)^{-{2 \over n\left(w_p(t) + 1\right)}}V_0^{-{2 \over n}}
\e^{-{2 \over w_p(t) + 1}\int^t \dot w_p(t')\ln a(t') dt'}\right\} \nn
&& + {\cal O}\left(f_1^2\right)\ .
\eea
 From Eqs.(\ref{FXXVII}) and (\ref{FXXVIII}),  the
expressions of
the entropy $S_p$, the extensive part of the energy $E_{pE}$ and the Casimir energy
$E_{pC}$ may be evaluated:
\bea
\label{WT6b}
S_p&=& f_{p0}\xi^{1 \over w_p(t)}\left(\left(1+{1 \over w_p(t)}\right) +
\left(1 + {1 \over w_p(t)} - {2 \over nw_p(t)}\right)f_{p1} \xi^{-{2 \over nw_p(t)}}\right)
\ , \nn
E_{pE}&=&{f_{p0} \over w_p(t)}a^{-nw_p(t)}V_0^{-w_p(t)}\xi^{1+{1 \over w_p(t)}}
\left(1 + \left(1 - {2 \over n} + w_p(t)\right)
f_{p1}\xi^{- {2 \over nw_p(t)}}\right) \ ,\nn
E_{pC}&=&-2f_{p0}f_{p1}a^{-nw_p(t)}V_0^{-w_p(t)}\xi^{1+{1 \over w_p(t)} - {2 \over nw(t)}}\ .
\eea
As $w_p(t)$ and $\xi$ are time-dependent, the entropy is not constant and not conserved.
Nevertheless, from (\ref{WT6b}) the Cardy-Verlinde
\cite{EV} like
formula ( a la Youm \cite{Youm})  (\ref{FXXVb}) is still valid:
\be
\label{WT7}
S_p \sim A_p\left[a^{nw_p} \sqrt{\left(2E_p-E_{pC}
\right)E_{pC}}\right]^{n \over (w_p+1)n-1} \ .
\ee
We should note, however, since
\be
\label{WT8}
A_p = f_{p0}\left(1+{1 \over w_p(t)}\right)\left(-{4f_{p0}^2 f_{p1} \over n}
\right)^{-{n \over 2\left(\left(w_p(t)+1\right)n - 1\right)}}
V_0^{wn \over (w_p(t)+1)n -1 }
\ee
and $w_p(t)$ depend on time, $A_p$ is not a constant but a function of the
time $t$. Thus, the entropy of the expanding universe with (negative)
time-dependent
equation of state is found.

Now, the FRW equations for the universe filled with matter with
pressure $p$ and  energy density $\rho$  are given by
\be
\label{II}
H^2={16\pi G \over n(n-1)}\rho - {k \over a^2}\ ,\quad
\dot H = -{8\pi G \over n-1}\left(\rho + p\right) + {k \over a^2}\ .
\ee
As in \cite{EV}, if we define the Hubble entropy $S_H$, the
Bekenstein-Hawking energy $E_{BH}$, and the Hawking temperature $T_H$ by
\be
\label{FXXXI}
S_H \equiv {(n-1)HV \over 4G}\ ,\quad
E_{BH}\equiv {n(n-1) V \over 8\pi G a^2}\ ,\quad
T_H\equiv - {\dot H \over 2\pi H}\ ,
\ee
the FRW equations can be rewritten in universal form  as
\bea
\label{FXXXII}
S_H&=& {2\pi a\over n}\sqrt{E_{BH}\left(2E - k E_{BH}\right)}\ ,\nn
kE_{BH}&=& n\left(E + pV - T_H S_H\right)\ ,
\eea
Furthermore with the Bekenstein entropy $S_B$ and the Bekenstein-Hawking
entropy
$S_{BH}$ as
\be
\label{BBH1}
S_B\equiv {2\pi a \over n}E\ ,\quad S_{BH}\equiv {(n-1)V \over 4Ga}\ ,
\ee
we obtain well-known relation between entropies
\be
\label{BBH2}
S_H^2 = 2 S_B S_{BH} - k S_{BH}^2\ .
\ee
In case of $k=1$, Eq.(\ref{BBH2}) can be rewritten as
\be
\label{BBH2b}
S_H^2 + \left(S_B - S_{BH}\right)^2 = S_B^2\ .
\ee
Then we find $S_H\leq S_B$. For the system with limited self-gravity, there
occurs the
Bekenstein bound \cite{bekenstein}:
\be
\label{BBH2c}
S\leq S_B\ .
\ee
This bound is useful for the case that the system has relatively low energy or
small volume. Then  Bekenstein entropy $S_B$ scales
as
$S_B \to \lambda^{1+{1 \over n}} S_B$ under the scale transformation
$V\to \lambda V$ and $E\to \lambda E$ \cite{EV}.

Eq.(\ref{FXXXII}) has a form similar to the second
equation in (\ref{FXIXc}) with
$w={1 \over n}$ and this equation is called  the cosmological Cardy-Verlinde formula.
The second equation in (\ref{FXXXI}) has a form similar to (\ref{FXVI}) and $E_{BH}$ may
correspond to the Casimir energy $E_C$. In \cite{EV}, the following
cosmological bound has been proposed:
\be
\label{FXXXIII}
E_C\leq E_{BH}\ .
\ee
As seen from the definition of $E_{BH}$ in (\ref{FXXXI}), we find
$E_{BH}\sim a^{n-2}$. If we consider phantom or quintessence as a matter field,
as seen from the last expression in (\ref{FXXX}), the behavior of the Casimir
energy is given by $E_C\sim a^{-nw}$. Then if
\be
\label{FXXXIV}
w < -1 + {2 \over n}
\ee
and $E_C$ is positive,
there is a critical radius $a_c$ where $E_C=E_{BH}$ and if the radius $a$ of the universe
is larger than the critical radius: $a>a_c$, the bound in (\ref{FXXXIII}) is
violated. Formally $a_c$ is given by
\be
\label{PP1}
a_c=\left[-{16\pi G f_{p0} f_{p1} V_0^{-w_p-1}C^{1 - {2 \over n}} \over n(n-1)}
\right]^{1 \over nw_p + n -2} \ ,
\ee
with the parameters $f_{p0}$, $f_{p1}$, and $C$, which may be determined
by
some initial conditions.
If we consider 4-dimensional spacetime ($n=3$), Eq.(\ref{FXXXIV}) gives
$w<-{1 \over 3}$, then for the quintessence $\left(-1<w<-{1 \over 3}\right)$, the cosmological
constant ($w=-1$), and the phantom ($w<-1$), there is always a critical radius $a_c$ and
the bound (\ref{FXXXIII}) is violated if $a>a_c$.

Similarly, one can discuss the entropy bounds for dark energy universe as in \cite{Youm}
even if $w_p$ depends on time. Although the entropy is not conserved,
the expression
of the entropy $S_p$  (\ref{WT7}) still holds. The quantity $\left(2E_p - E_{p}\right)E_{pC}$
inside the square root of (\ref{WT7}) has a maximum $E_p^2$ when $E_{pC}=E_p$. Then
\bea
\label{YYY1}
& S\leq A_p\left[a^{nw_p }E_p\right]^{n \over \left(w_p + 1\right) n - 1} \quad
& \mbox{for} \quad w_p> -1 + {1 \over n}\ ,\nn
& S\geq A_p\left[a^{nw_p} E_p\right]^{n \over \left(w_p + 1\right) n - 1} \quad
& \mbox{for} \quad w_p< -1 + {1 \over n}\ .
\eea
As $w_p$ depends on time, at some time, we may have $w_p> -1 + {1 \over n}$ and at
another time, $w_p< -1 + {1 \over n}$. If we define the Bekenstein entropy $S_{pB}$
for the dark energy as in (\ref{BBH1}): $S_{pB}\equiv {2\pi a \over n}E_p$, we find,
even if $w_p$ depends on time, the relation as in \cite{Youm}:
\bea
\label{YYY2}
& S\leq S_0 \left[a^{nw_p -1}S_B\right]^{n \over \left(w_p + 1\right) n - 1} \quad
& \mbox{for} \quad w_p> -1 + {1 \over n}\ ,\nn
& S\geq S_0 \left[a^{nw_p -1}S_B\right]^{n \over \left(w_p + 1\right) n - 1} \quad
& \mbox{for} \quad w_p< -1 + {1 \over n}\ .
\eea
Here $S_0$ is given by
\be
\label{YYY3}
S_0 = A_p \left({n \over 2\pi}\right)^{n \over \left(w_p + 1\right) n - 1} \ .
\ee
However, as $w_p$ and  $A_p$ depend on time, $S_0$ also depends on
time. If $w<-1<-1 + {1 \over n}$, the entropy can be negative (unphysical case) even if
the energy is positive. If $-1<w<{1 \over n}$, the entropy becomes negative
only when the
energy is negative.

\section{Entropy and energy in modified gravity}

In \cite{CDTT,CCT}, a gravitational alternative was suggested for the
dark energy
modifying the standard Einstein action  at low curvature by  $1/R$ term.
Such modified gravity may produce the current cosmic speed-up \cite{CDTT}
and may be naturally generated by string/M-theory \cite{NO}.
It represents some kind of higher derivative and non-local gravity,
and as such it may contain some instabilities \cite{C}.
Nevertheless, with some mild modifications at high curvature region
the theory is shown to be stable \cite{NO1} which is also supported
by quantum field theory \cite{NO1}.
Modified gravity was studied in Palatini form \cite{V},  and it seems that
it may be viable also in such a version.
Classically, its action may be mapped to an equivalent scalar-tensor theory.
We discuss below the
entropy, the energy and CV formula for accelerated universe in modified
gravity which provides the gravitational dark energy.

Let us start from  the
rather general 4-dimensional action:
\be
\label{RR1}
\hat S={1 \over \kappa^2}\int d^4 x \sqrt{-g} f(R)\ ,
\ee
where $\kappa^2=16\pi G$,
$R$ is the scalar curvature, and $f(R)$ is some arbitrary function.
By using the conformal transformation
\be
\label{RR7}
g_{\mu\nu}\to \e^\sigma g_{\mu\nu}\ ,
\ee
with
\be
\label{RR8b}
\sigma = -\ln f'(R)\ ,
\ee
etc., the action (\ref{RR1}) is rewritten as
\be
\label{RR10}
\hat S_E={1 \over \kappa^2}\int d^4 x \sqrt{-g} \left( R - {3 \over 2}g^{\rho\sigma}
\partial_\rho \sigma \partial_\sigma \sigma - V(\sigma)\right)\ .
\ee
Here
\be
\label{RR11}
V(\sigma)= \e^\sigma g\left(\e^{-\sigma}\right) - \e^{2\sigma} f\left(g\left(\e^{-\sigma}
\right)\right)= {A \over f'(A)} - {f(A) \over f'(A)^2}\ .
\ee
Here $g(B)$ is given by solving the equation $B=f'(A)$ with respect to $A$: $A=g(B)$ and
$A$ in (\ref{RR11}) is given by $A\equiv -\e^{2\sigma}$.
This is the standard form of the scalar-tensor theories where the scalar field
is fictitious \cite{NO1}.

We now consider the FRW cosmology in modified gravity.
 FRW metric in the physical (Jordan) frame is given by:
\be
\label{RR33}
ds^2 = - dt^2 + \hat a(t)^2 \sum_{i,j=1}^3 \gamma_{ij} dx^i dx^j\ .
\ee
The FRW equation in the Einstein frame has the following form:
\be
\label{RR40}
3 H_E^2 + {3k \over 2\hat{a_E}^2}={\kappa^2 \over 2}\left(\rho_{(\sigma E)}
+\rho_{(m)}\right)\ .
\ee
Here $\rho_{(m)}$ is the energy density of the matter but for simplicity, we neglect
the matter. We also concentrate on the $k=0$ case but the obtained results
 are
correct even for $k\neq 0$ case if the radius of the universe is large enough.
The Hubble constant $H_E$ in the Einstein frame is defined by
\be
\label{RR40b}
H_E\equiv {\dot{\hat{a_E}} \over \hat{a_E}}
\ee
with the scale factor $\hat{a_E}$ in the Einstein frame:
\be
\label{RR39}
\hat{a_E}=\e^{-{\sigma \over 2}}\hat a\ .
\ee
The contribution from the $\sigma$ field to the energy-momentum tensor $\rho_{(\sigma E)}$ is
given by
\be
\label{RR41}
\rho_{(\sigma E)}\equiv {1 \over \kappa^2}\left({3 \over 2}\dot \sigma^2 + V(\sigma)
\right)\ .
\ee
In the Einstein frame, the equation of motion for $\sigma$ has the following form:
\be
\label{RR43}
0=3\left(\ddot \sigma + 3 H_E \dot \sigma\right) + V'(\sigma) \ .
\ee
Assuming that when the curvature is small the action is given by
\be
\label{RR1b}
\hat S={1 \over \kappa^2}\int d^4 x \sqrt{-g} \left(R - {\tilde a \over R}\right)\ ,
\ee
the potential is given by
\be
\label{RR28}
V(\sigma) \sim {2 \over \sqrt{\tilde a}}\e^{{3 \over 2}\sigma}\ .
\ee
Since $\sigma = -\ln f'(R)\sim -\ln {\tilde a \over R^2}$,
$\sigma$ is negative and large.
Then the solution of equations (\ref{RR40}) and (\ref{RR43}) is given by
\be
\label{RR44}
\hat{a_E}=\hat{a_{E0}}\left({t_E \over t_0}\right)^{4 \over 3}\ ,
\quad \sigma = - {4 \over 3}\ln {t_E \over t_0}\ ,\quad
{t_0^2 \over \sqrt{\tilde a}}=4\ .
\ee
Here $t_E$ is the time coordinate in the Euclidean frame, which is related to
the time coordinate $t$ in the (physical) Jordan frame by $\e^{\sigma \over 2}dt_E
= dt$. As a result
\be
\label{RR45b}
3t_E^{1 \over 3}=t\ ,
\ee
and  in the physical (Jordan) frame the power law inflation occurs
\be
\label{RR45}
\hat a = \e^{{\sigma \over 2}} \hat{a_E} \propto t_E^{2 \over 3}\propto t^2\ ,
\ee
In general, if $p=w\rho$, the scale factor $a$ behaves as
\be
\label{RR46}
\hat a\sim t^{2 \over 3(w+1)}\ .
\ee
Then as we can see from (\ref{RR45}), in the Jordan frame we find $w=-{2 \over 3}$ and
from (\ref{RR45}), in the Einstein frame, $w=-2$. In fact,
 in the Einstein frame one has
\bea
\label{RRR1}
\rho_{(\sigma E)}&=& {1 \over \kappa^2}\left({3 \over 2}\dot \sigma^2 + V(\sigma)
\right) \sim {32 \over 3\kappa^2 t_E^2}\ ,\nn
p_{(\sigma E)}&\equiv& {1 \over \kappa^2}\left({3 \over 2}\dot \sigma^2 - V(\sigma)
\right) \sim -{16 \over 3\kappa^2 t_E^2}\ .
\eea
Although the Jordan frame is physical, as the separation to the gravity
and the matter
is more easy in the Einstein frame, we work in the Einstein frame for a
while.
 FRW equation (\ref{RR40}) can be rewritten in the form of the cosmological
CV formula with $n=3$ as
\be
\label{RRR2}
S_H^E= {2\pi a\over 3}\sqrt{E_{BH}^E\left(2E^E - k E_{BH}^E\right)}\ .
\ee
by defining
\be
\label{RRR3}
S_H^E \equiv {H_E V_E \over 2G}\ ,\quad E^E\equiv \rho_{(\sigma E)}V_E\ ,\quad
E_{BH}^E\equiv {3 V_E \over 4\pi G \hat a_E^2}\ ,\quad
V_E \equiv \hat a_E^3 \int d^3x \sqrt{-\gamma}\ ,
\ee
and $\kappa^2 = 16\pi G$.
The second FRW equation can be given by considering the derivative of the
(first) FRW equation (\ref{RR40}) with respective  Einstein time $t_E$
and can be
rewritten as
\be
\label{RRR4}
kE_{BH}^E= 3\left(E^E + p_{(\sigma E)}V_E - T_H^E S_H^E\right)\ .
\ee
Here
\be
\label{RRR5}
T_H^E\equiv - {1 \over 2\pi H_E}{d H_E \over dt_E}\ .
\ee
and we find
\be
\label{RRR5b}
p_{(\sigma E)}= - {1 \over 3H_E}{d\rho_{(\sigma E)} \over dt_E} - \rho_{(\sigma E)}\ .
\ee
In the physical Jordan frame, since $\hat a = \e^{{\sigma \over 2}} \hat{a_E}$ and
$\e^{\sigma \over 2}dt_E = dt$, the Hubble parameter is
\be
\label{RRR6}
H\equiv{1 \over \hat a}{d\hat a \over dt}={1 \over \hat a_E}{d\hat a_E \over dt_E}
{dt_E \over dt} + {1 \over 2}{d\sigma \over dt}
=H_E\e^{-{\sigma \over 2}} + {\dot \sigma \over 2}\ .
\ee
Then in the Jordan frame, the FRW equation can be rewritten as
\be
\label{RRR7}
3H^2 + {3k \over {\hat a}^2}= {\kappa^2 \over 2}\rho_{(\sigma)}\ ,\quad
\rho_{(\sigma)}\equiv\rho_{(\sigma E)}\e^{-\sigma} + H\dot \sigma - {{\dot \sigma}^2 \over 4}\ .
\ee
Defining
\be
\label{RRR8}
S_H \equiv {H V \over 2G}\ ,\quad E\equiv \rho_{(\sigma)}V\ ,\quad
E_{BH}\equiv {3 V \over 4\pi G \hat a^2}\ ,\quad
V \equiv \hat a^3 \int d^3 \sqrt{-\gamma}\ .
\ee
we obtain the cosmological Cardy-Verlinde formula:
\be
\label{RRR9}
S_H= {2\pi a\over 3}\sqrt{E_{BH}\left(2E - k E_{BH}\right)}\ .
\ee
By differentiating the FRW equation (\ref{RRR7}) with respect to $t$, one
gets
the second FRW equation:
\be
\label{RRR10}
{dH \over dt} - {k \over {\hat a}^2}={\kappa^2 \over 2}\left(\rho_{(\sigma)}
+ p_{(\sigma)}\right)\ ,\quad
p_{(\sigma)}\equiv  - {1 \over 3H}{d\rho_{(\sigma)} \over dt} - \rho_{(\sigma)}\ .
\ee
With the definition of the temperature $T_H$ by
\be
\label{RRR11}
T_H\equiv - {1 \over 2\pi H}{d H \over dt}\ ,
\ee
it follows
\be
\label{RRR12}
kE_{BH}= 3\left(E + p_{(\sigma)}V - T_H S_H\right)\ .
\ee
For the case of $k=0$, by substituting (\ref{RR44}), (\ref{RR45b}), and (\ref{RR45}) into
the expressions of $\rho_{(\sigma)}$ in (\ref{RRR7}) and $p_{(\sigma)}$ in (\ref{RRR10}),
we find
\be
\label{RRR13}
\rho_{(\sigma)}={\rho_{(\sigma)0} \over t^2}\ ,\quad
\rho_{(\sigma)0}\equiv {22 (27)^{2 \over 3} \over 3\kappa^2 t_0^{4 \over 3}} -12\ ,\quad
p_{(\sigma)}=-{2 \over 3}{\rho_{(\sigma)0} \over t^2}\ .
\ee
Eventually, it follows $w=-{2 \over 3}$ in the Jordan frame.

At the low temperature, as the field with lowest (negative) $w$ dominates,
we may have a equation similar to (\ref{FXXVb}) with $n=3$:
\bea
\label{EB1}
S &\sim& A_\sigma\left[a^{3w_\sigma} \sqrt{\left(2E_\sigma-E_{\sigma C}
\right)E_{\sigma C}}\right]^{3 \over 3w_\sigma+2} \ ,\nn
A_\sigma&\equiv& f_{\sigma0}\left(1+{1 \over w_\sigma}\right)\left(-{4f_{\sigma 0}^2
f_{\sigma 1} \over 3}
\right)^{-{3 \over 2\left(3w_\sigma+2\right)}}
V_0^{3w_\sigma \over 3w_\sigma + 2}\ .
\eea
Since $w_\sigma=-{2 \over 3}$, the exponents in (\ref{EB1}) diverges. Then in order that the
entropy is finite, the condition appears
\be
\label{EB2}
\left(-{4f_{\sigma 0}^2 f_{\sigma 1} \over 3}\right)^{-{3 \over 2}}V_0^{3w_\sigma}
\left[a^{3w_\sigma} \sqrt{\left(2E_\sigma-E_{\sigma C}\right)E_{\sigma C}}\right]^3=1\ .
\ee
We should also note that the solution  (\ref{RR44}) or (\ref{RR45}) is for $k=0$ case.
Then the Casimir force should vanish. In order to find the Casimir force, we need to consider
the $k\neq 0$ case. As the expansion over $k$ corresponds to the
expansion with respect to the
inverse of the radius of the universe, we may consider the perturbation with respect to $k$ in
order to obtain the Casimir energy.

We should also note that, as disucussed after (\ref{FXV}), since now $w$ is greater than $-1$
but negative, the entropy $S$ could be negative only if the energy is negative.


\section{Black hole thermodynamics}

We now consider the black hole solution in the modified gravity, whose
action is given by (\ref{RR1b}).
As it will be shown , its thermodynamical properties are also related to the
CV formula.
If we assume $R_{\mu\nu}\propto g_{\mu\nu}$, the
equation of motion is given by
\be
\label{RRR14}
0=\left(1+{\tilde a \over R^2}\right)R_{\mu\nu} - {1 \over 2}g_{\mu\nu}
\left(R - {\tilde a \over R}\right)\ ,
\ee
Then
\be
\label{RRR15}
R=\pm \sqrt{3\tilde{a}}\ ,\quad R_{\mu\nu}=\pm {\sqrt{3\tilde{a}} \over 4}g_{\mu\nu}\ .
\ee
A large class of solutions of (\ref{RRR15}) is given by the family of metrics
\bea
\label{RRR16}
&& ds^2 = -\e^{2\rho}dt^2 + \e^{-2\rho}dr^2 +
\sum_{i,j=1,2}g^{(2)}_{ij}dx^i dx^j \ ,\nn
&& \e^{2\rho}={1 \over r}\left(-\mu + k^{(2)} r - {\Lambda r^3 \over 3}\right)\ ,\quad
\Lambda=\mp\frac{\sqrt{3\tilde{a}}}{4}
\eea
embracing de Sitter (dS) and anti-de Sitter (AdS) black holes with any horizon topology.
Here $k^{(2)}$ is the Ricci curvature of the transverse manifold, as
given by the Ricci tensor $R^{(2)}_{ij}$ of the metric
$g^{(2)}_{ij}$, i.e. $R^{(2)}_{ij}=k^{(2)}g^{(2)}_{ij}$.
If $\Lambda<0$ ($\Lambda>0$), the spacetime is asymptotically anti-deSitter
(deSitter). In both cases the curvature radius will be defined by
$L^2=3/|\Lambda|=12/\sqrt{3\tilde{a}}$.


We shall mainly study the SAdS metric although our results apply
equally well to any horizon topology. The thermodynamical free energy can
be obtained
according to a quantum gravity tree-level formula involving the
Euclidean action $I_E$
\[
F(\beta)=\beta^{-1}I_E=\frac{\kappa}{2\pi}\,I_E
\]
where $\kappa$ is the surface gravity of the black
 hole. To pursue this program one has to regularize the volume
divergences. In anti-de Sitter gravity one can achieve this, essentially, by two well
known methods. One is the counterterm method inspired by the Maldacena
duality with conformal field theories, the other a background
subtraction chosen to correspond to the vacuum of the CFT. This
uniquely identifies it as anti-de Sitter space itself, with no matter
inside. The unregularized Euclidean action will be
\be
\label{eact}
I_E=-\frac{1}{16\pi G}\int\,d^4x\left(R-\frac{\tilde{a}}{R}\right)|g|^{1/2}-
\frac{1}{8\pi G}\oint\,K|h|^{1/2}d^3x
\ee
The Euclidean SAdS solution is given by (\ref{RRR16}) taking
$k^{(2)}=1$ and the metric $g^{(2)}_{ij}$ to be that of a round two-sphere
\be
ds^2=\left(1-\frac{\mu}{r}+\frac{r^2}{L^2}\right)
d\tau^2+\left(1-\frac{\mu}{r}+\frac{r^2}{L^2}\right)^{-1}dr^2+r^2d\omega_2^2
\label{esads}
\ee
where $d\omega_2^2$ is the line element of a $2$-sphere with unit
radius and volume $\omega_2=4\pi$. Moreover
$\tau\simeq\tau+\beta$ is periodically identified up to $\beta$ and
the curvature radius is  $L^2=12/\sqrt{3\tilde{a}}$.
This is a solution of (\ref{eact}) with $R=-\sqrt{3\tilde{a}}$.
Therefore it represents a spherically symmetric black hole
immersed in anti-de Sitter space.

The background metric will be (\ref{esads}) with $\mu=0$, i.e. anti-de
Sitter space at finite temperature $T=\kappa/2\pi$. This has zero
gravitational entropy, since there is no horizon. The action
(\ref{eact}) for the metric (\ref{esads}) is easily seen to be
\be
I_E=\frac{\sqrt{\tilde{a}}\beta}{
3\sqrt{12}G}(R_m^3-r_+^3)+ \mbox{``boundary terms''}
\label{a1}
\ee
where $R_m$ is an upper bound for the radial integration and $r_+$ is
the radius of the horizon. The action of the background is
\be
I_{EB}=\frac{\sqrt{\tilde{a}}\beta}{3
\sqrt{12}G}R_0^3+ \mbox{``background boundary terms''}
\label{aB}
\ee
where again $R_0$ is a radial cutoff. Now a meaningful comparison of
the black hole free energy with the vacuum free energy (empty AdS
space) requires that the vacuum metric on the surface $r=R_0$ be
asymptotically coincident with the actual metric on the surface
$r=R_m$. This matching condition ensures that the boundary temperatures
in the black hole and the background, be equal. A simple check gives
the matching condition that, asymptotically
\[
R_0=R_m-\frac{\mu L^2}{6R_m^2}
\]
Using this into (\ref{aB}) and subtracting the result from (\ref{a1}),
gives the regularized action\footnote{One finds that the boundary
terms do not contribute to the final result.}
\be
\Delta I_E=\frac{\sqrt{\tilde{a}}\beta}{6\sqrt{12}G}(\mu L^2-2r_+^3)
\label{rega}
\ee
We note that the mass parameter $\mu$ and $r_+$ are functions of $\beta$
through the defining relations
\be
\label{emme}
\mu=r_++\frac{r_+^3}{L^2}
\ee
\be\label{beta}
\beta=\frac{4\pi L^2r_+}{\mu L^2+2r_+^3}
\ee
Hence the entropy could be computed by the familiar thermodynamical relation
\[
S=\beta\partial_{\beta}\Delta I_E-\Delta I_E
\]
Instead we may use an easier way. We note that both $R-\tilde{a}/R$ as
well as $R-2\Lambda$ are proportional to $\sqrt{\tilde{a}}$, so $I_E$ must be
proportional to the action as computed in Einstein gravity. Denoting
this as $I_{AdS}$, a simple computation gives
\be
I_E=\frac{4}{3}\,I_{AdS}
\ee
We know that the entropy in Einstein gravity is $A/4G$, so we
immediately conclude that in $1/R$ gravity the entropy must be
\be
\label{entropy}
S=\frac{4}{3}\,\frac{A}{4G}=\frac{A}{3G}
\ee
So black holes in modified gravity are a little bit more entropic than
expected.
We may confirm this result by using the Noether charge method.
In this case the formula is\cite{wald}
\be
S=4\pi\int_{S^2}\frac{\partial\cal{L}}{\partial R}\,d^2x
\ee
where ${\cal L}={\cal L}(R)$ is the Lagrangian density and the integral is
over the horizon at $r=r_+$. In our case
${\cal L}=\sqrt{g}(R-\tilde{a}/R)/16\pi G$, so
\[
S=\frac{A}{4G}\left(1+\frac{\tilde{a}}{R^2}\right)=\frac{4}{3}\,\frac{A}{4G}
\]
as a simple computation will confirm using $R^2=3\tilde{a}$.  These
calculations can be done in any spacetime dimensions, say $d$. Then
(\ref{entropy}) generalizes to
\be
\label{entropy2}
S=\frac{2d}{d+2}\,\frac{A}{4G}
\ee
Note that for the black hole with the size of FRW universe, the entropy
is
defined by the Bekenstein-Hawking entropy $S_{BH}$  (\ref{BBH1}). Then
the above result
(\ref{entropy}) and (\ref{entropy2}) indicates that $S_{BH}$ should also
be modified by
the factor ${2d \over d+2}$ if compared with the FRW universe in Einstein
gravity.

The higher entropy of black hole in $1/R$-gravity means that they
are more massive than in Einstein theory, since by the first law
$dM=TdS$. The precise prediction should just be that $M$ is larger by the
factor $z=2d/(d+2)$.

An asymptotically $SAdS_d$ black hole in General Relativity has an
excitation energy over the AdS vacuum which can be computed by
canonical methods, by means of the formula
\be
\label{massf}
M=-\frac{1}{8\pi G}\oint\,N(\Theta-\Theta_0)\sqrt{\sigma}d^{d-2}x
\ee
Here we integrate over a $(d-2)$-dimensional sphere at infinity,
contained in a Cauchy surface of equal time, the lapse function
$N=\sqrt{-g_{tt}}$, times the trace of the second fundamental form of
the sphere as embedded in the Cauchy surface, after a regularizing
subtraction from empty AdS space. For the metric (\ref{esads}) one finds
\be
M=\frac{(d-2)\omega_{d-2}}{16\pi G}\,\mu
\ee
This can be expressed as a function of the black hole radius by using
the condition $N(r_+)=0$, which is
\be
\label{emme1}
\mu=r_+^{d-3}+\frac{r_+^{d-1}}{L^2}
\ee
In theories with an AdS dual, this relation can be interpreted as the
energy of a CFT living on the boundary  of
AdS spacetime, and leads to a CV formula for AdS black holes. In higher
derivatives gravity, and this is just our case, things may be not so
straightforward. For a theory whose Lagrangian $L=L(R)$ is a function of
the scalar curvature, the above mass can be related to a Noether
charge\cite{wald1} which is proportional to $\partial L/\partial R$,
as in the entropy derivation given above. More than this, it is this
Noether charge that enters the formulation of the first law for
stationary black holes in diffeomorphism covariant theories of
gravity\cite{wald,wald1}. The result is the mass formula (\ref{massf}),
except that the integrand gets multiplied with $16\pi G\partial
  L/\partial R$ evaluated on the background solution, where
$L=(R-\tilde{a}/R)/16\pi G$ is the actual Lagrangian. This gives all
  masses an extra coefficient
\[
1+\frac{\tilde{a}}{R^2}=\frac{4}{3}
\]
It is therefore clear that the Cardy-Verlinde formula for AdS black
holes\cite{cai,edi}, being the square root of a quadratic function of
all the relevant energies, will give the entropy the $4/3$ coefficient
too, in accord with our calculations.

\section{Hydrodynamical examples testing the holographic entropy bound}

The suggestion of Kovtun {\it et al.} \cite{kovtun03} that there may exist in cosmology
a universal lower bound on $\eta/s$ - $\eta$ being the shear viscosity and $s$ the entropy
content per unit volume - is interesting, since it may be of fundamental importance.
These authors are concerned with the infrared properties of theories whose gravity duals
contain a black brane with a nonvanishing Hawking temperature, the point being that the
infrared behavior is governed by hydrodynamical laws. If we for definiteness consider a stack
of $N$ non-extremal D3 branes in type IIB supergravity, the metric near the horizon is given by
\begin{equation}
ds^2= \frac{r^2}{R^2}[-f(r)dt^2+dx^2+dy^2+dz^2]+ \frac{R^2}{r^2f(r)}dr^2+R^2d\Omega_5^2,
\label{1}
\end{equation}
where $R\propto N^{1/4}$ is a constant, and $f(r)=1-r_0^4/r^4$ with $r_0$ being the horizon.
The Hawking temperature of this metric is $T=r_0/\pi R^2$, and $\eta$ and $s$ are given by
\begin{equation}
\eta=\frac{1}{8}\pi N^2T^3, \quad s=\frac{1}{2}\pi^2N^2T^3.
\label{2}
\end{equation}
Thus, in dimensional notation
\begin{equation}
\frac{\eta}{s}=\frac{\hbar}{4\pi k_B}=6.08 \times 10^{-13}~{\rm K\,s}.
\label{3}
\end{equation}
The conjecture of Kovtun {\it et al.} (see also \cite{BL}) is that the
value in Eq.~(\ref{3}) is a {\it lower bound}
for $\eta/s$. Since this bound does not involve the speed of light, the authors even conjecture
that this bound exists for all systems, including non-relativistic ones.

The idea has recently been further elaborated in \cite{karch03}, arguing
that the
bound follows from the generalized covariant entropy bound. From Eq.(\ref{BBH2}), there is the
Bekenstein (and also the holographic) entropy bound, which is used to prove the new bound to
shear viscosity.

The purpose of this section is to elucidate this holographic idea by considering some examples
explicitly. We will choose examples from general physics. Our scope is thus
 wider than in the previous sections; our  aim is to investigate the generality
 of the entropy bound.
We will consider three examples, the first  taken
from ordinary hydrodynamics, the second from the  theory of the universe in the
beginning of its plasma era, and finally the third taken from the  very early universe under
 conditions corresponding to the  Kasner metric. The third example is
presumably the one of main interest; the shear viscosity concept is after all a concept that
relates to a physical situation that is {\it anisotropic}. Moreover, we will discuss
the validity of the Cardy-Verlinde entropy formula in the case
of viscous cosmology, thus elaborating on the previous treatment on this topic in \cite{brevik02}.

The central inequality that we intend to analyze, is thus
\begin{equation}
\frac{\eta/s}{\hbar/4\pi k_B} > 1.
\label{4}
\end{equation}

\noindent
{\it Example 1. Hydrodynamics: Small Reynolds number flow.}
The following setup taken from  ordinary hydrodynamics involves both the shear
viscosity $\eta$ and the entropy density $s$: Assume that a solid sphere with radius $R$ and
with high thermal conductivity $\lambda$ is immersed in a uniform flow passing it at small
Reynolds numbers. We take the origin in the center of the sphere, and use spherical coordinates
with the polar axis in the direction of the undisturbed velocity ${\bf u}$ of the stream.
The equation of thermal conduction is
\begin{equation}
\nabla^2 T=-\frac{\eta}{2\lambda}(v_{i,k}+v_{k,i})^2,
\label{5}
\end{equation}
where ${\bf v}$ is the fluid velocity for $r\geq R$. Inserting Stokes' formula (applicable
at low Reynolds numbers) for ${\bf v}$, the solution for the temperature distribution $T(r)$
can be written as \cite{landau87}
\[
 T(r)-T_0=\frac{9u^2\eta}{4\lambda}
\Bigg\{ \left( \frac{3}{4}\frac{R^2}{r^2}-\frac{5}{3}\frac{R^3}{r^3}
+\frac{R^4}{r^4}-\frac{1}{12}
\frac{R^6}{r^6}\right) \cos^2\theta
\]
\begin{equation}
+\frac{2}{3}\frac{R}{r}-\frac{3}{4}\frac{R^2}{r^2}+\frac{5}{9}\frac{R^3}{r^3}-
\frac{1}{6}\frac{R^4}{r^4}-\frac{1}{36}\frac{R^6}{r^6} \Bigg\},
\label{6}
\end{equation}
where $T_0$ is the constant reference temperature at infinity. The boundary conditions are
$T=T_1=$ const and $\int (\partial T/\partial r)r^2\sin \theta \,d\theta=0$ for $r=R$. From
Eq.~(\ref{6}) it is seen that $\Delta T\equiv T_1-T_0=5u^2\eta/8\lambda$.

One may ask: What is the appropriate value to be inserted for the entropy density $s$?
Taking water as an example, one might use the handbook value for $s$, resulting in
$\eta/s=2.3\times 10^{-10}~{\rm K\,s}$, as in Ref.~\cite{karch03}. However, in our opinion
the physically most natural value to use for $s$ in the present example is the one
associated with the temperature difference $\Delta T$. This amounts to setting
\begin{equation}
s=\rho c_p \int_{T_0}^{T_1}\frac{dT}{T} \simeq \rho c_p\frac{\Delta T}{T_0},
\label{7}
\end{equation}
$c_p$ being the specific heat capacity at constant pressure. We then get
\begin{equation}
\frac{\eta}{s}=\frac{8\nu T_0}{5u^2}\frac{1}{Pr},
\label{8}
\end{equation}
where $\nu=\eta/\rho$ is the kinematic viscosity and $Pr=\nu \rho c_p/\lambda$ the Prandtl
number. We choose the moderate velocity $u=1$ mm/s to keep the Reynolds number small, and
take $T=300$ K. Then, with $\nu=0.010\, {\rm cm^2/s},~ Pr= 6.75$ \cite{landau87} we get
\begin{equation}
\frac{\eta}{s}=71\, {\rm K\;s}
\label{9}
\end{equation}
as a typical value. The inequality (\ref{4}) is obviously satisfied.

\noindent
{\it Example 2. Plasma era in the early universe.}
As the next step  we consider the initial stage of the the plasma era in the
early universe. This can be taken to occur at about $t=1000$ s after the big bang, when
the universe is characterized by ionized H and He in approximate equilibrium with radiation
(cf., for instance, \cite{harrison73,borner88,brevik94,brevik97}). The number densities
of electrons and photons are equal, $n \simeq 10^{19}\; {\rm cm^{-3}}$, the temperature is
$T\simeq 4\times 10^8$ K, and the energy density is $\rho c^2=a_rT^4$, where
$a_r=\pi^2 k_B^4/(15\hbar^3c^3)=7.56\times 10^{-15}\; {\rm erg\,cm^{-3}\,K^{-4}}$ is
the radiation constant. The pressure is $p=\rho c^2/3$. The presence of energy dissipation and
viscosity coefficients in the cosmic fluid is due to the fact that the thermal equilibrium is
not quite perfect. From relativistic kinetic theory one can calculate the viscosity
coefficients. Let $x=m_ec^2/k_BT$ be the ratio between electron rest mass and thermal energy;
when $x \gg 1$ it is convenient to use the polynomial approximations
 \cite{caderni77} (cf. also \cite{brevik94}) for the evaluation of the shear viscosity
$\eta$ and the bulk viscosity $\zeta$:
\begin{equation}
\eta=\frac{5m_e^6 \,c^8\zeta(3)}{9\pi^3\hbar^3 \,e^4 \,n}x^{-4},
\label{10}
\end{equation}
\begin{equation}
\zeta=\frac{\pi c^2\hbar^3 n}{256\, e^4\zeta(3)}x^3,
\label{11}
\end{equation}
$\zeta(3) =1.202$ being the Riemann zeta function. At $T=4\times 10^8$ K one has $x=14.8$,
leading to
\begin{equation}
\eta=2.8\times 10^{14}\; {\rm g\,cm^{-1}\,s^{-1}},\quad
\zeta=7.0\times 10^{-3}\;{\rm g\,cm^{-1}\,s^{-1}}.
\label{12}
\end{equation}
We note that both $\eta$ and $\zeta$ now contain $\hbar$, and also that $\eta$ is enormously
larger than $\zeta$.

The entropy density, in view of the radiation dominance, is given by
\begin{equation}
s=\frac{4}{3}a_rT^3=6.45\times 10^{11}\; {\rm erg\,cm^{-3}\,K^{-1}},
\label{13}
\end{equation}
and so
\begin{equation}
\frac{\eta}{s}=435\; {\rm K\,s}.
\label{14}
\end{equation}
This value is surprisingly enough of the same order of magnitude as  the value given
in Eq.~(\ref{9}). There seems to be no simple reason why this should be so; the physical
conditions in the two cases are widely different.

So far, we assumed a radiation dominated FRW universe. What
happens if the universe is instead filled with matter obeying the
relation $p=w \rho c^2$, with $w$ constant and negative? To
investigate this point let us go back to Eq.~(\ref{FXIXb}), in
which the sub-extensive parts are neglected. For the ratio $s/\rho
c^2$, where $s=S/V$ and $\rho c^2 = E/V$, we obtain
\begin{equation}
\frac{s}{\rho c^2}=\frac{1+w}{T} . \label{14a}
\end{equation}
This expression is seen to be independent of the prefactor $f_0$.
Let us assume that the energy density at $T=4\times 10^8\,$K is the same as before, i.e.,
$\rho c^2=a_rT^4=1.94\times 10^{20}\; {\rm erg\,cm^{-3}}$. Then $s$ is found from (\ref{14a}),
and taking the shear viscosity to be given by (\ref{11}) as before, we obtain the following
simple equation
\begin{equation}
\frac{\eta}{s}=\frac{578}{1+w}.
\label{14b}
\end{equation}
We see that except in the case where $w$ is close to  $-1$, the order of magnitude of $\eta/s$
is roughly the same as above. It is moreover evident that the expression (\ref{14b}) is
physically meaningful only when $w >-1$ (the viscosity $\eta$ has always to be positive, for
general thermodynamical reasons). We thus see that the inclusion of shear viscosity implies
that it is only the case of quintessence that is of physical interest. The case of phantoms,
$w<- 1$, leads to negative entropies and is in the present context excluded.

\noindent
{\it Example 3. The Kasner universe.}
Our third  example is taken from the theory of the very early
universe. From ordinary hydrodynamics we know that the shear viscosity comes into play whenever
there are fluid sheets sliding with respect to each other. Correspondingly, in a relativistic
formulation, the most natural circumstances under which  $\eta$ is expected to be of
significance are when anisotropy is brought into consideration. It becomes natural
to focus attention on the anisotropic Kasner metric
\begin{equation}
ds^2=-dt^2+t^{2p_1}dx^2+t^{2p_2}dy^2+t^{2p_3}dz^2,
\label{15}
\end{equation}
where the numbers $p_1,p_2,p_3$ are constants. The two numbers $P$
and $Q$ are defined by
\begin{equation}
P=\sum_{i=1}^3 p_i,\quad Q=\sum_{i=1}^3 p_i^2.
\label{16}
\end{equation}
In a vacuum Kasner space, $P=Q=1$. Here, we assume that there is an isotropic fluid with
energy density $\rho$ and pressure $p$ immersed in this space. Both $\rho$ and $p$, as well
as the viscosity coefficients $\eta$ and $\zeta$, are assumed to be dependent on time but
independent of position.
If $U^\mu=(U^0,U^i)$ is the fluid's four-velocity, the energy-momentum tensor is
\begin{equation}
T_{\mu\nu}=\rho U_\mu U_\nu+(p-\zeta \theta)h_{\mu\nu}-2\eta \sigma_{\mu\nu},
\label{17}
\end{equation}
where $h_{\mu\nu}=g_{\mu\nu}+U_\mu U_\nu$ is the projection tensor, $\theta={U^\mu}_{;\mu}$
is the scalar expansion, $\theta_{\mu\nu}=\frac{1}{2}(U_{\mu;\alpha}h_\nu^\alpha
+U_{\nu;\alpha}h_\mu^\alpha)$ is the expansion tensor, and $\sigma_{\mu\nu}
=\theta_{\mu\nu}-\frac{1}{3}h_{\mu\nu}\theta$ is the shear tensor.

Consider now the Einstein equations, taking the cosmological constant $\Lambda$ to be zero.
With $\kappa^2=16\pi G$ we obtain from
$R_{\mu\nu}=\frac{1}{2}\kappa^2 (T_{\mu\nu}-\frac{1}{2}g_{\mu\nu}T_\alpha^\alpha)$ the
two equations
\begin{equation}
P-Q+\frac{3}{4}\kappa^2 t\zeta P=\frac{1}{4}\kappa^2 t^2(\rho+3p),
\label{18}
\end{equation}
\begin{equation}
p_i(1-P-\kappa^2 t\eta)+\frac{1}{4}\kappa^2 t(\zeta+\frac{4}{3}\eta)P
=-\frac{1}{4}\kappa^2 t^2(\rho-p).
\label{19}
\end{equation}
The structure of the Einstein equations leads to the time relationships
\begin{equation}
\rho(t)=\rho_0(t_0/t)^2, \quad p(t)=p_0(t_0/t)^2,
\label{20}
\end{equation}
\begin{equation}
\zeta(t)=\zeta_0\, t_0/t, \quad \eta(t)=\eta_0\, t_0/t,
\label{21}
\end{equation}
where $\{\rho_0,p_0,\zeta_0,\eta_0\}$ refer to the chosen initial instant $t=t_0$.
We can then write the equations such that they contain time-independent quantities only:
\begin{equation}
P-Q+\frac{3}{4}\kappa^2 \zeta_0t_0P=\frac{1}{4}\kappa^2 \,t_0^2(\rho_0+3p_0),
\label{22}
\end{equation}
\begin{equation}
p_i(1-P-\kappa^2 \eta_0 t_0)+\frac{1}{4}\kappa^2 t_0(\zeta_0+\frac{4}{3}\eta_0)P
=-\frac{1}{4}\kappa^2 \,t_0^2(\rho_0-p_0).
\label{23}
\end{equation}
Let us consider the production of entropy. First, for the Bianchi type-I spaces
the average expansion anisotropy parameter $A$ is defined as \cite{gron85}
\begin{equation}
A=\frac{1}{3}\sum_{i=1}^3\left( 1-\frac{H_i}{H}\right)^2,
\label{24}
\end{equation}
where $H_i=\dot{a}_i/a_i$ with $a_i=t^{p_i}$ are the directional Hubble factors
and $H=\frac{1}{3}\sum_1^3 H_i$ is the average Hubble factor. Accordingly, in our case
\begin{equation}
A=\frac{3Q}{P^2}-1.
\label{25}
\end{equation}
Next, the entropy current four-vector is $S^\mu=n\sigma U^\mu$, where $n$ is the baryon
number density and $\sigma=s/n$ the nondimensional entropy per baryon. In general,
\begin{equation}
{S^\mu}_{;\mu}=\frac{2\eta}{T}\sigma_{\mu\nu}\sigma^{\mu\nu}+\frac{\zeta}{T}\theta^2,
\label{26}
\end{equation}
meaning in the comoving frame of reference ($\dot{\sigma}=d\sigma/dt$)
\begin{equation}
\dot{\sigma}=\frac{3P^2}{nTt^2}\left(\zeta+\frac{2}{3}\eta A \right).
\label{27}
\end{equation}
As we would expect, the anisotropy in general provides a significant contribution to
the growth of entropy, in view of the large magnitude of $\eta$. However, let us go back
to Eq.~(\ref{23}): this equation tells us that all the $p_i$ have to be {\it equal} in the
present case. With $p_1=p_2=p_3 \equiv b$ we get for the {\it isotropic} Kasner space
\begin{equation}
b=\frac{1}{6}\left[ 1+\frac{3}{4}\kappa^2 t_0 \zeta_0+\sqrt{\left( 1
+\frac{3}{4}\kappa^2 t_0\zeta_0\right)^2+3\kappa^2 t_0^2\,(\rho_0-p_0)}\right].
\label{28}
\end{equation}
It is seen that the shear viscosity is absent. Equation (\ref{27}) reduces to
\begin{equation}
\dot{\sigma}=\frac{3P^2}{nk_B Tt^2}\zeta,
\label{29}
\end{equation}
when written in dimensional form.

Let us evaluate the expression (\ref{29}). Due to the proportionality to the small bulk
viscosity we can insert for $n$ and $T$ as if the cosmic fluid were ideal. Thus from
conservation of particle number, $n\propto a^{-3}$, and from conservation of entropy,
$a\propto T^{-1}$. As moreover $t\propto T^{-2}$, we can write Eq.~(\ref{29}) as
\begin{equation}
\dot{\sigma}=\frac{3P^2\zeta_0}{n_0 k_B T_0 t_0}\frac{1}{t}.
\label{30}
\end{equation}
Thus $\sigma -\sigma_0 \propto \ln (t/t_0)$ is the increase in specific entropy when $t$
increases from $t_0$ to $t$. Multiplying with the particle density $n$ we obtain an
expression for the corresponding increase $s-s_0$ in entropy density. Recalling the
expression in Eq.~(\ref{21}) for $\eta$ we then derive as our main result the following
expression for the sought ratio:
\begin{equation}
\frac{\eta}{s}=\frac{\eta_0t_0}{s_0t}\left[1+
\frac{3P^2\zeta_0}{s_0 T_0 t_0}\left(\frac{t_0}{t}\right)^{3/2}\ln \frac{t}{t_0}\right]^{-1}.
\label{31}
\end{equation}
It is of interest to evaluate this expression at $t=t_0$. Let us identify $t_0$ with the
instant at which $T=10^{12}$ K, i.e., at $t=2\times 10^{-4}$ s.  Then $n_0=6\times 10^{29}\;
{\rm cm^{-3}},\, \rho_0=4.5\times 10^{34}\; {\rm erg\, cm^{-3}}$. This temperature is a kind
of limit for standard cosmological theory. If $T>10^{12}$ K the universe consists of many
kinds of particles and antiparticles, but when $T$ has fallen below this value the large
number of hadrons has disappeared, and the universe consists of leptons, antileptons, photons,
and nucleons. We then have \cite{caderni77}
\begin{equation}
\eta=\frac{3\pi c\hbar^4}{608\,m_e \,G_F^2}x, \quad
\zeta= \frac{\pi c \hbar^4}{7776 \,m_e\,G_F^2}x^5,
\label{32}
\end{equation}
which is valid when $x=m_ec^2/k_BT$ is small. Here, the weak coupling constant is given
by $G_Fc/\hbar^3=10^{-5}m_p^{-2}$.

With $T_0=10^{12}$ K we get $x=5.94\times10^{-3}$, and so we have at this instant
\begin{equation}
\eta_0=1.8\times 10^{23}~ {\rm g\,cm^{-1}\,s^{-1}}, \quad
\zeta_0=6.0\times 10^{12}~{\rm g\,cm^{-1}\,s^{-1}}.
\label{33}
\end{equation}
The entropy density is calculated approximatively by assuming radiation dominance, such
as before. Then, from from $s=\frac{4}{3}a_rT^3$ we get
\begin{equation}
s_0=1.0\times 10^{22}~{\rm erg\,cm^{-3}\,K^{-1}}.
\label{34}
\end{equation}
Thus,
\begin{equation}
\frac{\eta_0}{s_0}=18~{\rm K\,s}.
\label{35}
\end{equation}
Once again, we end up with the same order-of-magnitude result for the ratio $\eta/s$
as before, when we choose to work at the instant $t=t_0$. However, Eq.~(\ref{31}) tells us
that $\eta/s$ {\it diminishes} with increasing $t$, and approaches zero
when $t\rightarrow \infty$. It means that $\eta/s$ cannot in this case be subject to
a lower bound. The Kasner case thus provides a counterexample to the
 suggestion in Eq.~(\ref{4}). Of course, this can be considered as rather
academical as current universe is not the anisotropic one.

Actually, it follows already from the thermodynamical formalism  that the lower bound
in Eq.~(\ref{4}) cannot
be universal. At least this is so in a phenomenological theory, in which $\eta$ and $\zeta$
are arbitrary input parameters. Namely, from Eqs.~(\ref{26}) or (\ref{27}) it is seen that
the specific entropy  rate of change involves both $\eta$ and $\zeta$. Let us imagine
that $\eta$ is kept constant while $\zeta$ is changing. Therewith $\dot{\sigma}$, and
accordingly $\sigma$ itself, as well as the ratio $\eta/s$, change. If this ratio were
subject to a lower bound, this would correspond to the existence of a maximum value of $s$.
However, we may make $\sigma$ and $s$ as large as we wish, by inserting sufficiently large
value of $\zeta$ in Eq.~(\ref{27}). Recall in this context the way in which viscosity
coefficients are introduced in fluid mechanics: they are based on the assumption that
{\it first order} velocity gradients are sufficient to construct the contribution to the  stress tensor due to deviations from thermal equilibrium. The theory is thus approximate already from the outset.

 The discussion of Verlinde \cite{EV} about the
holographic bound on the sub-extensive entropy associated with the Casimir energy, assumed
a radiation dominated FRW universe. As shown in \cite{brevik02}, the same entropy formula
holds if the fluid possesses a constant, though small, bulk viscosity. Similarly,
the generalized entropy formula \cite{Youm} for the case that
the state equation is $p=w\rho$ with $w$ a constant (still assuming a FRW metric),
was also found to hold in the presence of the same kind of viscosity \cite{brevik02a}.

One may ask: How does the  entropy formula look if the
cosmic fluid possesses both a shear viscosity and a bulk viscosity? The answer is
immediate,  if the anisotropy is originally introduced via the Kasner metric. As shown above
the  Einstein equations wash out the anisotropies,  and we are left
with an isotropic Kasner metric whose  scale factor is  $t^b$, where $b$ is  given
by Eq.~(\ref{28}). The anisotropy factor $A$  vanishes, and the production of entropy is
governed by the bulk viscosity $\zeta$; cf. Eq.~(\ref{27}).

Let us assume that $\zeta$ is constant and small, so that  we can adopt the same expression
for $a(t)$ as in the case of a nonviscous fluid.  The argument can be given similarly to
that given in \cite{brevik02a}:  Taking $n=3$ we see that the quantity $\rho a^{3(w+1)}$ can
be considered as a function of $n\sigma$. Since $E \sim \rho a^3$ and $S \sim n\sigma a^3$,
it follows that $Ea^{3w}$ is independent of $V$ and is a function of $S$ only.
The conventional decomposition of the total energy $E(S,V)$ into an extensive part and
a sub-extensive part, $E(S,V)=E_E(S,V)+\frac{1}{2}E_C(S,V)$, together with the scale
transformations $E_E(\lambda S,\lambda V)=\lambda E_E(S,V),\,
E_C(\lambda S,\lambda V)=\lambda^{1/3}E_C(S,V)$, leads to
\begin{equation}
E_E=\frac{\alpha}{4\pi a^{3w}}S^{w+1},\quad E_C=\frac{\beta}{2\pi a^{3w}}S^{w+1/3},
\label{36}
\end{equation}
where $\alpha$ and $\beta$ are constants. We thus get, when we reinstate $a=t^b$,
\begin{equation}
S=\left[ \frac{2\pi t^{3bw}}{\sqrt{\alpha \beta}}\sqrt{(2E-E_C)E_C}\right]^{\frac{3}{3w+2}}.
\label{37}
\end{equation}
This is the Kasner-induced form of our previous expression (\ref{FXIXc}), when $n=3$.
It will be of interest to understand better the connection between CV
formula and shear viscosity bound. However, this requires the non-trivial
generalization of CV formula for anisotropic universe with shear
viscosity.

\section{Discussion}

 In summary, we studied the entropy of FRW universe filled with dark
energy and its representation in the form of holographic CV formula.
This investigation shows that the expression of the entropy in terms
of energy and Casimir energy depends on the equation of state in a quite
complicated form. It is only for a radiation dominated FRW universe
the corresponding CV formula acquires the form typical for 2d CFT entropy.
On the same time, for negative or time-dependent equation of state
such a formula seems to have nothing to do with 2d CFT, being still
related with holography. Nevetheless, there
exists another, cosmological
CV formula which is very useful to derive the entropy bounds and which is
the same for any type of matter under consideration. It is remarkable
that universality of cosmological CV formula together with the fact that
it predicted by the form of FRW equations proves its holographical
origin. Of course, the actual reasons for such a manifestation of the
holographic principle in the modern universe remain to be obscure.
(Some hints maybe drawn from brane-world approach.)
Furthermore, all above conclusions remain to be true in the modified
gravity which is considered as gravitational alternative for dark energy.
This should not seem strange after all as modified gravity maybe
re-written in the classically equivalent form as kind of scalar-tensor
gravity with matter described by scalar field.

The black hole thermodynamics in modified gravity is also considered.
The black hole entropy law is slightly different (by numerical
factor) from the standard case of the Einstein gravity.
In the last section we analyze the recently proposed bound for ratio
of shear viscosity with entropy density. This bound seems to follow from
the
Bekenstein entropy bound. As shear viscosity is absent in the current
isotropic universe, we concentrate on the early universe at plasma era
or anisotropic Kasner universe where newly proposed bound seems to be
violated.

The important lesson drawn from this and other studies of the entropy
of FRW universe is that holographic principle does not distinguish
whether dark energy is present or not. For instance, the cosmological CV
formula
is the same whatever is the equation of state. This indicates that the
origin
of dark energy should be searched within fundamental theory,
perhaps within string/M-theory.

\section*{Acknowledgments}

We are grateful to J.E. Lidsey for participation at the early stage of
this work.
This investigation has been supported in part by the Ministry of
Education, Science, Sports and Culture of Japan under the grant n.13135208
(S.N.), by BFM2003-00620 project (S.D.O.), by RFBR grant 03-01-00105
(S.D.O.), by LRSS grant 1252.2003.2
(S.D.O.) and by the Program INFN(Italy)-CICYT(Spain).

\end{document}